\documentclass[camera,letterpaper,nomarginnotes,nonarrowgutter]{jpaper}
\usepackage{amsmath,amssymb,amsfonts}
\usepackage{algorithmic}
\usepackage{graphicx}
\usepackage{textcomp}
\usepackage{xcolor}
\usepackage{fancyhdr}
\usepackage{hyperref}
\usepackage{tabularx}
\usepackage{multirow}
\usepackage{colortbl}
\usepackage{tikz}
\usepackage{makecell}
\usepackage{enumitem}
\usepackage{booktabs}
\usepackage{colortbl}
\usepackage{hhline}
\usepackage{subfig}
\usepackage{graphicx,scalerel}
\usepackage{textgreek}

\makeatletter
\g@addto@macro{\normalsize}{%
 \setlength{\abovedisplayskip}{2pt plus 1pt minus 1pt}
 \setlength{\belowdisplayskip}{1pt plus 1pt minus 1pt}
  \setlength{\abovedisplayshortskip}{1pt}
  \setlength{\belowdisplayshortskip}{1pt}
  \setlength{\intextsep}{1pt plus 1pt minus 1pt}
  \setlength{\textfloatsep}{2pt plus 1pt minus 1pt}
  \setlength{\skip\footins}{3pt plus 1pt minus 1pt}}
 \titlespacing\section{2pt}{2pt plus 2pt minus 2pt}{2pt plus 2pt minus 2pt}
 \titlespacing\subsection{2pt}{2pt plus 2pt minus 2pt}{2pt plus 2pt minus 2pt}
\makeatother

\definecolor{dmchecked}{HTML}{000000}

\definecolor{dmchecked2}{HTML}{000000}
\newcommand{\checkednew}[1]{\textcolor{dmchecked2}{#1}}

\definecolor{dmgreen}{HTML}{000000}

\definecolor{akblue}{HTML}{000000}
\newcommand{\AK}[1]{\textcolor{akblue}{#1}}

\begin{document}
\title{\textit{throttLL'eM}: Predictive GPU Throttling for Energy Efficient LLM Inference Serving}

\author{
    Andreas Kosmas Kakolyris\textsuperscript{$\dagger,\ddagger$}\qquad
    Dimosthenis Masouros\textsuperscript{$\ddagger$}\qquad
    Petros Vavaroutsos\textsuperscript{$\ddagger$}\\
    Sotirios Xydis\textsuperscript{$\ddagger$}\qquad
    Dimitrios Soudris\textsuperscript{$\ddagger$}
    \vspace{0mm}\\\\
    ETH Z{\"u}rich\textsuperscript{$\dagger$}\qquad
    National Technical University of Athens\textsuperscript{$\ddagger$}
}

\maketitle

\renewcommand{\headrulewidth}{0pt}
\fancypagestyle{firstpage}{
\fancyhead{} 
\fancyhead[C]{
} 
\renewcommand{\footrulewidth}{0pt}
}
\thispagestyle{firstpage}

\begin{abstract}
As Large Language Models (LLMs) gain traction, their reliance on power-hungry GPUs places ever-increasing energy demands, raising environmental and monetary concerns.
Inference dominates LLM workloads, presenting a critical challenge for providers: minimizing energy costs under Service-Level Objectives (SLOs) that ensure optimal user experience. 
In this paper, we present \textit{throttLL'eM}, a framework that reduces energy consumption while meeting SLOs through the use of instance and GPU frequency scaling. 
\textit{throttLL'eM} features mechanisms that project future Key-Value (KV) cache usage and batch size. Leveraging a Machine-Learning (ML) model that receives these projections as inputs, \textit{throttLL'eM} manages performance at the iteration level to satisfy SLOs with reduced frequencies and instance sizes. We show that the proposed ML model achieves $R^2$ scores greater than 0.97 and miss-predicts performance by less than 1 iteration per second on average. Experimental results on LLM inference traces show that \textit{throttLL'eM} achieves up to 43.8\% lower energy consumption and an energy efficiency improvement of at least $1.71\times$ under SLOs, when compared to NVIDIA's Triton server. \textit{throttLL'eM} is publicly available at \href{https://github.com/WilliamBlaskowicz/throttLL-eM}{\textit{https://github.com/WilliamBlaskowicz/throttLL-eM}}.

\end{abstract}

\section{Introduction}
\label{sec:introduction}

Large Language Models (LLMs) have surged in popularity over the last couple of years, driven by advancements in deep learning and highlighted by the remarkable success of ChatGPT from OpenAI~\cite{achiam2023gpt}. 
This rapid adoption, combined with the reliance of LLMs on high-end GPUs with significant power and energy consumption has raised concerns about their environmental impact.~\cite{samsi2023words,patel2023polca}.
Recent studies indicate that inference dominates the computational workload, accounting for over 90\% of the overall LLM compute cycles within data centers\cite{patterson2022carbon,patel2024characterizing}. While the environmental impact of data centers is primarily attributed to hardware manufacturing~\cite{chasingCarbon}, their operational footprint is also significant; the energy required for a single GPT-4 response can reach $\approx3$Wh~\cite{energ-per-query}. 
As userbases for LLM services grow into the millions~\cite{chatgpt_stats}, their energy footprint is projected to become substantial, potentially reaching 1050 TWh by 2026~\cite{iea-electricity}, rivaling entire countries.

Minimizing the energy footprint of LLM inference serving systems is crucial not only for environmental reasons but also from the providers' perspective.
Efficient utilization of resources allows data center providers to reduce their operating costs, while also facilitating compliance to power budget constraints, enforced by contracts with utility companies~\cite{zhang2021flex}.
However, reducing energy consumption often conflicts with the need for high performance.
LLM-based services (e.g., ChatGPT) are highly interactive, as users submit their queries (e.g., questions) and await for ``on-the-fly'' responses.
This user expectation of low latency is often quantified through Service-Level-Objectives (SLO), e.g., maintaining the $99^{th}$ percentile of response latency under a specific threshold~\cite{patel2024splitwise}.

Designing energy-efficient LLM serving systems that operate under SLOs is challenging due to their inherent unpredictability compared to traditional neural networks.
We identify three major barriers that complicate this task:
\begin{enumerate}[leftmargin=*]
    \item \textbf{Autoregressive nature of LLMs:} Unlike traditional feed-forward neural networks (e.g., CNNs) with fixed execution steps, LLMs exhibit inherent stochasticity in their inference process. Response generation proceeds iteratively, processing one token (word or sub-word unit) at a time until an End-of-Sequence (\texttt{EOS}) token is generated or a predefined maximum token limit is reached.
    This dynamic token generation length makes it difficult to precisely pre-allocate computational resources for each inference request.
    \item \textbf{Dynamic Batch Composition:} Techniques like \textit{inflight batching}~\cite{gyeong2022orca} leverage the autoregressive nature of LLMs to improve throughput by allowing requests to enter and exit a batch dynamically. 
    However, as shown in Sec.~\ref{subsec:motivation-heatmaps}, different batch sizes introduce significant performance variability, possibly increasing time-between-tokens and end-to-end latency by up to 45\%, thus complicating energy optimization under SLOs.
    \item\textbf{Variable Memory Footprint:} LLMs exhibit variable memory footprint due to their autoregressive nature. 
    Within each iteration, memory is dynamically reserved for the growing sequence length and only deallocated once a request is completed.
    This takes the form of a variably sized Key-Value (KV) cache, which stores the information associated with each token. While techniques like \textit{paged attention}~\cite{kwon2023pagedattention} have mitigated memory fragmentation issues, managing the KV cache remains challenging. Notably, increased KV cache usage can lead to an up to 18.2\% performance degradation in the inference process (cf. Sec.~\ref{subsec:motivation-kv-cache-analysis}).
\end{enumerate}

Managing the energy consumption of inference serving systems through traditional power-capping~\cite{li2020thunderbolt} and power-oversubscription~\cite{patel2023polca,kumbhare2021prediction} techniques falls short in the case of LLMs.
These solutions often operate in a static manner, leading to high tail latencies under load~\cite{MarkInferenceServing}, potentially conflicting with the specified SLOs set by providers~\cite{samsi2023words}.
Moreover, techniques designed for energy optimization of traditional neural network inference serving systems~\cite{ye2023deep}, such as \textit{delayed}~\cite{crankshaw2017clipper} or \textit{coordinated}\cite{nabavinejad2022coordinated} batching, cannot be directly applied to LLMs due to their unpredictability.
Finally, race-to-idle strategies~\cite{kim2015racing,le2011slow} are also rendered inefficient since LLM inference serving systems typically experience a continuous stream of requests~\cite{patel2024splitwise,wang2024burstgptrealworldworkloaddataset}, making it challenging to achieve true idle states for extended durations.

Given these limitations, dynamic GPU frequency scaling (throttling) forms a promising approach for optimizing energy efficiency in LLM inference serving.
Unlike static power management techniques, frequency scaling offers a dynamic control knob, allowing the system to adjust its power consumption to the real-time workload demands of incoming inference requests. 
Crucially, frequency scaling can be applied at the granularity of individual LLM iterations (which typically span milliseconds), whereas traditional techniques operate on the scale of entire queries (which span several seconds). 
This fine-grained, iteration-level control allows throttling to be applied in a more granular manner, thus greatly enhancing energy efficiency while meeting predefined SLOs.

However, effectively applying frequency scaling to LLM inference remains challenging due to the inherent unpredictability in the text generation process exhibited by these models.
While recent works address generation length prediction~\cite{jin2023s3, qiu2024efficient, hu2024inference, NEURIPS2023_ce7ff340}, there is a notable gap in modeling the impact of LLM-intrinsic characteristics and leveraging these insights for energy-efficient LLM inference serving.

This paper presents \textit{throttLL'eM}, an energy-efficient LLM inference serving framework. 
By leveraging dynamic GPU frequency scaling, \textit{throttLL'eM} minimizes energy consumption while adhering to latency and throughput SLOs. 
The system relies on a projection mechanism that estimates KV cache utilization and batch size and a performance prediction model that forecasts system throughput at future LLM iterations. 
These predictions guide a throttling mechanism, which identifies the minimum frequency that meets target SLOs, thereby optimizing energy usage. 
\textit{throttLL'eM} also features an autoscaling mechanism that adjusts the engine's parallelism level based on the incoming workload.
Our contributions are:

\setlist[itemize]{topsep=0pt, partopsep=0pt, parsep=0pt, itemsep=0pt}
\begin{itemize}[leftmargin=*]
\item \textbf{Comprehensive analysis:} We conduct a thorough analysis of GPU frequency, KV cache usage, batch size, and instance parallelism, providing key insights into LLM performance and energy efficiency.

\item \textbf{Predictive modeling:} We develop an analytical model that forecasts KV cache usage and batch size for future LLM iterations with \textit{mean absolute prediction errors} of 2.26\% and 0.19\%, respectively. Additionally, we introduce a machine learning model that predicts iteration-level LLM performance with $R^2$ scores exceeding 0.97.

\item \textbf{\textit{throttLL'eM} framework:} We introduce \textit{throttLL'eM}, a framework that leverages the aforementioned models for instance autoscaling and GPU frequency throttling, optimizing LLM inference energy consumption while meeting SLOs.
\end{itemize}

\noindent \textit{throttLL'eM} reduces energy consumption by an average of 24.7\% without autoscaling, reaching 43.8\% with autoscaling enabled, when compared to the state-of-the-art Triton inference server\cite{triton-inference-server}. The proposed framework targets a p$99^{th}$ response time SLO equal to that of the baseline system running at peak load and a TBT SLO matching the human reading rate. \textit{throttLL'eM} is publicly available at \href{https://github.com/WilliamBlaskowicz/throttLL-eM}{\textit{https://github.com/WilliamBlaskowicz/throttLL-eM}}.

\section{Background on LLM Inference}
\label{sec:background}

\begin{figure}[t]
\centering
\subfloat[Inference serving SW/HW stack.]{
    \includegraphics[width=0.23\linewidth,keepaspectratio=true]{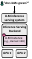}
\label{fig:llm-hwsw-stack}}
\subfloat[LLM auto-regressivity. During \textit{prefill}, the KV cache is generated. During \textit{generation} it is updated.]{
    \includegraphics[width=0.695\linewidth,keepaspectratio=true]{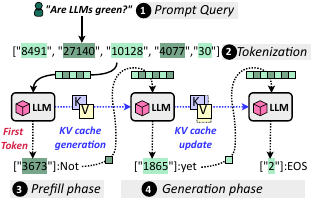}
\label{fig:llm-text-generation}}
\caption{LLM \textbf{a}) serving stack and \textbf{b}) text generation process.}
\label{fig:llm-background}
\end{figure}

LLMs are neural networks with billions of parameters that heavily rely on the Transformer architecture~\cite{vaswani2017attention}.
Figure~\ref{fig:llm-hwsw-stack} shows an overview of a typical LLM inference serving stack.
Inference servers (e.g., Triton~\cite{triton-inference-server}, LMDeploy~\cite{2023lmdeploy}) bridge the gap between users and pre-trained LLMs.
They act as a frontend, receiving user queries and interacting with inference serving backends (e.g., TensorRT-LLM backend~\cite{tensorRT-LLM-backend}, vLLM~\cite{kwon2023pagedattention}), which handle the deployment of LLM instances on the underlying infrastructure.

\begin{figure*}[t]
\centering
\subfloat[] {
\includegraphics[width=0.4\columnwidth]{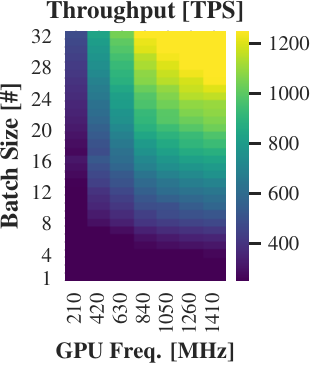}
    \label{fig:motivation-throughput}
}
\hspace{\fill}
\subfloat[]{
    \includegraphics[width=0.33\columnwidth]{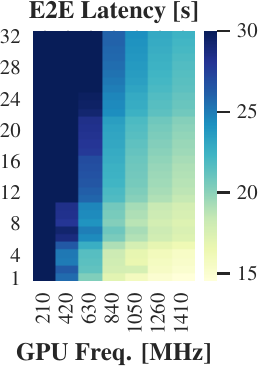}
    \label{fig:motivation-e2e}
}
\hspace{\fill}
\subfloat[] {
    \includegraphics[width=0.33\columnwidth]{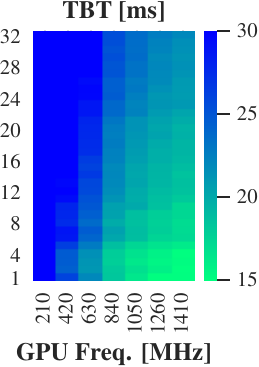}
    \label{fig:motivation-tbt}
}
\hspace{\fill}
\subfloat[] {
    \includegraphics[width=0.34\columnwidth]{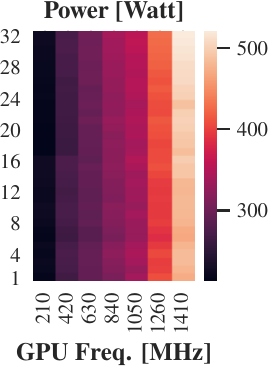}
    \label{fig:motivation-power}
}
\hspace{\fill}
\subfloat[] {
    \includegraphics[width=0.31\columnwidth]{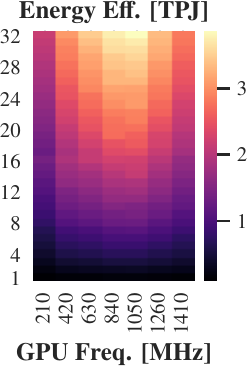}
    \label{fig:motivation-nrgeff}
}
\caption{Impact of different batch sizes and GPU frequencies on \textbf{a}) system throughput, \textbf{b}) end-to-end (E2E) latency, \textbf{c}) time between tokens (TBT), \textbf{d}) GPU power consumption and \textbf{e}) energy efficiency.}\label{fig:cost}
\end{figure*}

\noindent\textbf{LLM text generation process:} LLMs generate text in a step-by-step manner, one token (word or sub-word unit) at a time, as shown in Fig~\ref{fig:llm-text-generation}.
The process begins with a prompt from the user, which is tokenized before being fed into the model.
The tokenization vector is passed to the LLM, generating the \textit{first token}. 
This is known as the \textit{prefill} or \textit{prompt} phase and is compute-bound~\cite{patel2024splitwise}.
The LLM then enters an iterative loop until it generates an End-of-Sequence (EOS) token or reaches a maximum length limit (\texttt{\small max\_tokens}). 
During each iteration, the LLM considers the prompt and previously generated tokens to calculate token occurrence probabilities and select a new one. 
The generated token is appended to the sequence, and the process repeats. 
This phase is referred to as \textit{decoding} or \textit{generation} and is known to be memory bound~\cite{agrawal2023sarathi}.

\noindent\textbf{Key-Value (KV) cache:} During the \textit{prefill} phase the input tokens are processed, generating their initial representations (values) stored with their corresponding token identities (keys) in the KV cache.
These key-value pairs are high-dimensional tensors, containing vectors for each attention head of the Transformer.
For the decoder-only Transformer architecture used by many LLMs, these vectors remain identical across iterations and can thus be cached.
In the generation phase, the LLM iteratively predicts tokens, checking the KV cache for existing representations. 
If found, the cached representation is used; otherwise, a new representation is computed and stored.

\noindent\textbf{Inflight Batching:} \textit{Inflight batching}~\cite{gyeong2022orca} is an optimization technique which capitalizes on the autoregressive nature of LLMs, by allowing new requests to be added to or removed from an existing batch while it is being processed.

\noindent\textbf{Instance Parallelism:} Modern LLM serving frameworks~\cite{tensorRT-LLM-backend,kwon2023pagedattention} support partitioning (sharding), enabling model distribution among many GPUs. 
This allows for executing large models on GPUs that would otherwise not be able to store LLM weights individually.
Three key approaches exist to distribute the workload across multiple GPUs~\cite{nvidia-parallelisms}. 
\textit{Distributed Data parallelism (DDP)} creates identical copies of the model across multiple GPUs.
\textit{Tensor parallelism (TP)} splits the model's weight tensors across multiple GPUs, enabling parallel computation for specific operations.
Finally, \textit{Pipeline parallelism (PP)} assigns consecutive LLM layers or segments to different GPUs, similar to instruction pipelining in CPUs.

\noindent\textbf{LLM inference performance metrics:} Several metrics have been proposed to evaluate LLM performance and serve as potential SLOs\cite{patel2024splitwise,stojkovic2024towards}. \textit{Time To First Token} (TTFT) measures the responsiveness of the inference engine, i.e., the time to generate the first token after query submission. Throughput metrics include \textit{Time Between Tokens} (TBT), \textit{LLM iterations per second} (IPS), and \textit{tokens per second} (TPS). \textit{End To End latency} (E2E) represents the total time from query submission to completion. 
To quantify energy efficiency, we consider \textit{tokens per Joule} (TPJ), measuring the number of tokens generated per unit of energy expended.

\section{Energy Efficient LLM Inference Serving: Challenges \& Opportunities}
\label{sec:motivation}

In this section we examine the impact of various factors that affect LLM performance and energy efficiency.
We perform this analysis on a NVIDIA DGX A100 system, utilizing the Llama2-13B model as a representative example of LLMs \footnote{Since most LLMs are built upon a backbone of stacked Transformer blocks, similar conclusions can be drawn for other models.}, with a tensor parallelism of two, unless otherwise stated.

\subsection{Performance-Energy Trade-offs in LLM Inference}
We investigate the interplay between performance and energy efficiency in LLM inference, aiming to identify the optimal operating points that align with specific performance metrics. 
Our analysis focuses on the following parameters: i) batch size, indicating the number of requests currently scheduled on the GPU; ii) GPU frequency, which directly impacts the latency of various LLM kernels; and iii) KV cache usage, representing the volume of data read by these kernels.

\noindent\subsubsection{Impact of Batch Size and GPU Frequency} 
\label{subsec:motivation-heatmaps}
We first examine the impact of batch size and GPU frequency on the performance and power consumption of LLM inference. To this end, we spawn batches of different sizes, ranging from 1 to 32 requests wide.
To avoid any variation or bias due to the context of the queries, all queries are identical with their prompt and generation lengths fixed to a specific value of 1 input token and 1024 generation tokens. 
Figures \ref{fig:motivation-throughput}-\ref{fig:motivation-nrgeff} show the average throughput, E2E latency, TBT, power consumption, and energy efficiency during the lifetime of each batch. 

\noindent\textbf{Performance Analysis: }Regarding performance related metrics (Fig.\ref{fig:motivation-throughput}-\ref{fig:motivation-tbt}), we first observe that throughput increases with batch size and GPU frequency, as shown in Figure \ref{fig:motivation-throughput}. 
Higher GPU frequencies enable faster processing of tokens, while larger batch sizes increase parallelism, contributing to higher overall throughput (TPS). \AK{Notably, when changing from a batch size of 1 at 210 MHz to a batch size of 32 at 1410 MHz, throughput experiences a threefold increase.} 
This trend partially aligns with those observed for the E2E latency and TBT metrics.
Figures \ref{fig:motivation-e2e} and \ref{fig:motivation-tbt} show that both E2E latency and TBT generally decrease with higher GPU frequencies due to reduced computation times. 
On the other hand, both metrics worsen as batch size increases, indicating that larger batch sizes incur overheads due to increased concurrent parallel processing and resource contention. \AK{Overall, both E2E and TBT approximately double when moving from the ``high frequency, low batch size'' to the ``low frequency, high batch size'' area.}

\noindent\textbf{Power Analysis:} Figure~\ref{fig:motivation-power} shows the power consumption for different GPU frequencies and batch sizes. 
As expected, power increases with higher GPU frequencies due to increased switching activity and operating voltage required to sustain higher processing speeds. We note a greater than twofold increase in power draw between the lowest and highest available GPU frequencies.
Interestingly, power consumption remains relatively constant across different batch sizes for a given GPU frequency, indicating that power draw is primarily influenced by the GPU's operating frequency rather than workload size.
This uniformity of power consumption across batch sizes, combined with the variability in performance metrics, suggests that the system's energy efficiency can be optimized by carefully selecting the appropriate GPU frequency for the desired batch size, thereby balancing power draw and performance.

\noindent\textbf{Energy Efficiency Analysis:} To further support the above statement, Fig.~\ref{fig:motivation-nrgeff} presents the energy efficiency of the system, i.e., the number of tokens generated per Joule.
We see that processing larger batches improves energy efficiency for all frequencies.
\AK{Counterintuitively to Fig.~\ref{fig:motivation-power}, reductions in frequency do not always correspond to increased efficiency.} We observe a clear sweet spot: efficiency increases with lower GPU frequencies up to a point (1050 MHz) before dropping again.
Notably, using the maximum batch size with a lower, more efficient 1050 MHz frequency can yield a significant 37.4\% boost in energy efficiency with minimal impact on performance (-6.25\%, +8.26\%, +5.41\% throughput, E2E latency, and TBT respectively). For low frequencies ($<$ 840 MHz) the reduction in power is offset by lower throughput, diminishing overall efficiency while simultaneously worsening performance.
The above observations suggest that frequency scaling offers a valuable control knob for energy-efficient LLM serving. 
By dynamically adjusting the GPU frequency based on workload size (batch size) and aiming for the sweet spot where efficiency is maximized, we can achieve significant energy savings without sacrificing substantial performance.
We note that optimal GPU frequency is architecture- and LLM-dependent and may vary with different generation lengths.

\noindent$\star$ \textit{\textbf{Takeaways:} Batch size significantly impacts both performance and energy efficiency, with larger batches improving throughput and efficiency but worsening E2E latency and TBT.
Low GPU frequencies are neither performant nor energy-efficient.
The frequency of optimal energy efficiency differs from that of peak performance, highlighting frequency scaling as a promising mechanism for exploring this trade-off.}

\subsection{Implications of KV cache usage on LLM inference}
\label{subsec:motivation-kv-cache-analysis} 

\begin{figure}[t]
\centering
\hspace{\fill}
\subfloat[KV vs. Throughput for different batch sizes.]{
    \includegraphics[width=0.29\columnwidth]{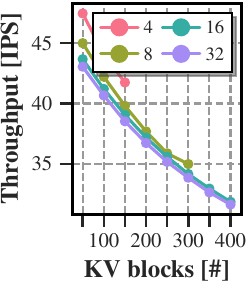}
    \label{fig:motivation-KV-Throughput-Batch}
}
\hspace{\fill}
\subfloat[KV vs. TBT for different batch sizes.] {
\includegraphics[width=0.29\columnwidth]{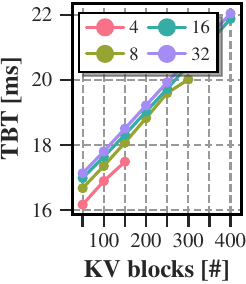}
    \label{fig:motivation-KV-TBT-Batch}
}
\hspace{\fill}
\subfloat[KV vs. Power for different frequencies.] {
\includegraphics[width=0.3\columnwidth]{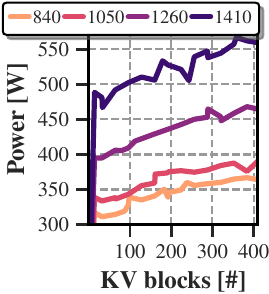}
    \label{fig:motivation-KV-Power-Freq}
}
\hspace{\fill}

\subfloat[Timeline of KV cache utilization, TBT and Throughput. Batch size is kept equal to 32 and GPU frequency is set to maximum.]{
    \includegraphics[width=.95\columnwidth]{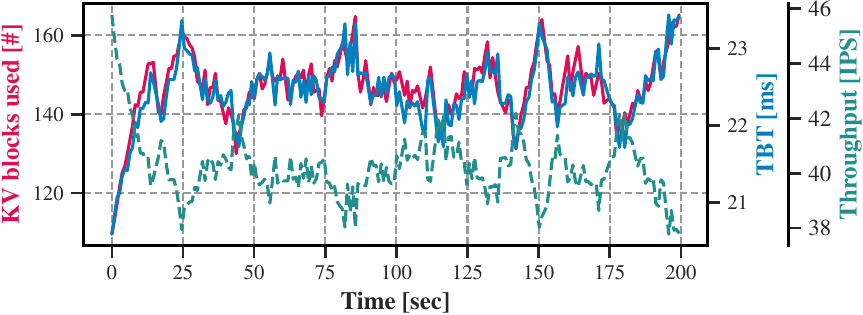}
    \label{fig:motivation-KV-TBT-Thr-Time}
}

\caption{Analysis on the impact and correlation of KV cache utilization to different performance metrics.}\label{fig:kv-implications}
\end{figure}

\noindent\textbf{Impact of KV on performance:} Figures~\ref{fig:motivation-KV-Throughput-Batch} and \ref{fig:motivation-KV-TBT-Batch} show the impact of KV cache utilization on throughput (IPS) and TBT for different batch sizes as the token generation process progresses and the number of KV blocks allocated in the GPU's DRAM increases. 
The results reveal that both the number of allocated KV blocks and the batch size affect performance. 
We observe that as the number of blocks increases there is clear degradation in Throughput and a linear increase in TBT.
This trend is consistent across all batch sizes.
While this trend is similar across all batch sizes, smaller batches achieve better performance for the same number of allocated KV blocks, due to the fewer tokens that need to be generated at each iteration, resulting in reduced resource contention within the GPU. 

\noindent\textbf{Impact of KV on power consumption:}
Figure~\ref{fig:motivation-KV-Power-Freq} shows the power consumption for a batch of 32 queries as the KV block count increases (reflecting the generation process) under different frequency levels.
A positive correlation between KV blocks and power consumption is observed across all frequency levels, which is attributed to the increased number of DRAM reads per iteration.
Notably, the power consumption curve steepens at higher frequencies as the number of KV blocks grows, indicating a more rapid power increase per additional KV block.
This variation in \AK{gradient} suggests that careful consideration of KV cache management and frequency scaling can lead to optimized energy efficiency.

\noindent\textbf{KV/Performance correlation analysis: } Our analysis has demonstrated a clear relationship between KV cache utilization, throughput, and TBT. 
To further examine this correlation, we conduct an inference experiment for 200 seconds while maintaining constant batch size (32) by introducing new requests of random generation lengths upon completion of existing ones, all under maximum frequency.
Figure~\ref{fig:motivation-KV-TBT-Thr-Time} shows that KV usage and TBT curves overlap, indicating a direct relationship, while throughput appears to be a mirrored image of KV cache usage.
Statistical analysis yields a Pearson correlation of 0.92 between KV and TBT and -0.92 between KV and Throughput, indicating that KV usage can serve as an accurate proxy for modelling performance. Repeating the experiment at lower batch sizes increases the Pearson correlation, due to less interference from newly arrived requests entering the prefill phase and stalling the token generation process.

\noindent$\star$ \textit{\textbf{Takeaways:} KV cache utilization significantly impacts both performance and power consumption. 
The high correlation between KV utilization and performance-related SLOs emphasizes its effectiveness as a proxy for performance modeling.}

\subsection{LLM partitioning implications on performance and energy}
\label{subsec:motivation-parallelism}
\begin{figure}[t]
\centering
\hspace{\fill}
\subfloat[Throughput vs. batch size]{
    \includegraphics[width=0.42\columnwidth]{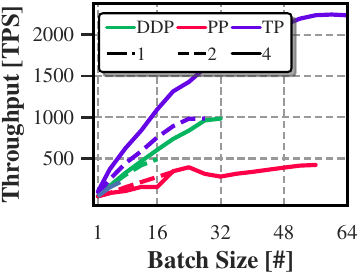}
    \label{fig:motivation-engine-size-throughput}
}
\hspace{\fill}
\subfloat[Efficiency vs. batch size]{
    \includegraphics[width=0.399\columnwidth]{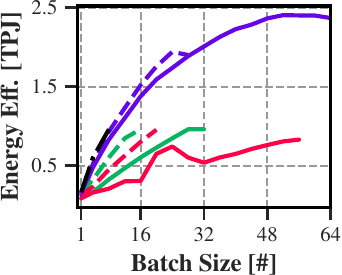}
    \label{fig:motivation-engine-size-efficiency}
}
\hspace{\fill}

\caption{Impact of different LLM partitioning approaches on \textbf{a}) throughput  and \textbf{b}) energy efficiency.
}\label{fig:motivation-parallelism-implications}
\end{figure}
As described in Sec.~\ref{sec:background}, LLMs support various techniques for partitioning the model across multiple GPUs, each impacting system efficiency differently.
Figures~\ref{fig:motivation-engine-size-throughput} and \ref{fig:motivation-engine-size-efficiency} show the throughput (TPS) and energy efficiency (TPJ) respectively, for different partitioning techniques (DDP, PP, TP) and parallelism levels across various batch sizes on a single node.
Regarding throughput, TP consistently outperforms DDP/PP, exhibiting a throughput improvement of $1.54/2.74\times$ and $1.79/6.26\times$ for parallelism levels 2 and 4 respectively, at the maximum batch size supported by all configurations.
Notably, TP also supports larger attainable batch sizes, demonstrating its ability to handle more concurrent requests and provide increased serving capacity for an equal number of GPUs. 
As expected, increasing the level of parallelism enhances performance due to the increased computational resources and memory capacity provided by additional GPUs, which results in more available KV cache blocks, allowing the system to process a greater number of requests simultaneously.
Efficiency-wise, TP again outperforms DDP and PP. 
However, smaller engine sizes provide better energy efficiency up to the maximum batch size they can support. 
Specifically, we see that TP2 achieves up to 9.66\% higher TPJ compared to TP4, when running close to its maximum batch size, indicating that smaller engines should be prioritized whenever possible to optimize energy efficiency.

\noindent$\star$ \textit{\textbf{Takeaways:} For intra-node scaling, TP configuration outperforms DDP and PP in both performance and energy efficiency. 
The trade-off between performance and energy efficiency at different levels of parallelism underscores the need for dynamic engine sizing based on real-time workload characteristics.
}

\subsection{Unpredictability in LLM inference serving systems}
\label{subsec:motivation-trace-analysis}
Inference serving systems exhibit significant unpredictability due to the diversity of input prompts and workload traffic variations~\cite{patel2024splitwise,wang2024burstgptrealworldworkloaddataset}. 
To illustrate this, we analyze a 60-minute production trace from Azure LLM inference services \cite{patel2024splitwise}.

\begin{figure}[t]
\centering

\hspace{\fill}
\subfloat[Distribution of prompt (top) and generated tokens (bottom).]{
    \includegraphics[width=0.43\columnwidth]{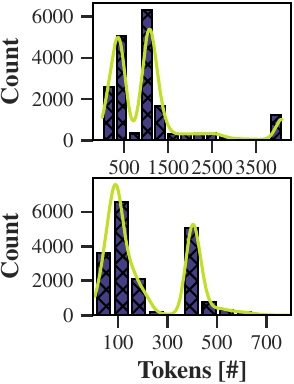}
    \label{subfig:azure-trace-analysis-queries}
}
\hspace{\fill}
\subfloat[Histogram of requests (top) and min and max RPS per bin (bottom).]{
    \includegraphics[width=0.43\columnwidth]{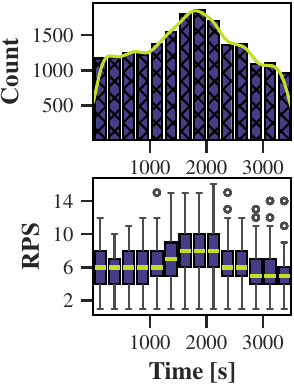}
    \label{subfig:azure-trace-analysis-rps}
}
\hspace{\fill}

\caption{Analysis of Azure's LLM inference trace~\cite{patel2024splitwise}.}
\label{fig:azure-trace-analysis}
\end{figure}

\noindent\textbf{Length Distribution:} Figure~\ref{subfig:azure-trace-analysis-queries} shows the distribution of prompt tokens (top) and generated tokens (bottom).
Query prompts can reach up to 4000 tokens, with most queries falling between 0-1500 tokens.
Similarly, generated tokens range from 10 to 700, with the majority between 100 and 400 tokens.
The long-tail distributions observed for both prompt and generated tokens underscore the necessity for dynamic resource allocation strategies to handle the varying computational demands of different queries.
Moreover, the unpredictable nature of token generation lengths further complicates resource management due to its direct impact on KV cache utilization and overall system performance (as shown in Sec.~\ref{subsec:motivation-kv-cache-analysis}).

\noindent\textbf{Arrival Pattern:} Figure~\ref{subfig:azure-trace-analysis-rps} shows the distribution of requests (top) and RPS (bottom) over the 60-minute trace, segmented into 4-minute bins.
The distribution exhibits non-uniformity, with a peak around the midpoint of the trace, indicating periods of lower and higher activity.
Additionally, the request arrival rate shows high variability, with median values ranging between 5 and 8 RPS and the distribution extending up to 16 RPS.
The minimum RPS per bin is one, indicating a continuous workload.
Given that the generation phase typically spans multiple seconds, the system experiences sustained activity without idle periods, rendering race-to-idle techniques inefficient for energy optimization.

\noindent$\star$ \textit{\textbf{Takeaways:} The long tails of token length distributions and high variability in arrival patterns necessitate dynamic resource allocation for optimal management. LLM inference systems rarely reach idle states, rendering race-to-idle techniques incapable of providing optimal energy efficiency.}

\section{\textit{throttLL'eM}: SLO-aware \& Energy-Efficient LLM Inference Serving}
\label{sec:methodology}

\begin{figure}[t]
\centering   
        \includegraphics[width=\columnwidth]{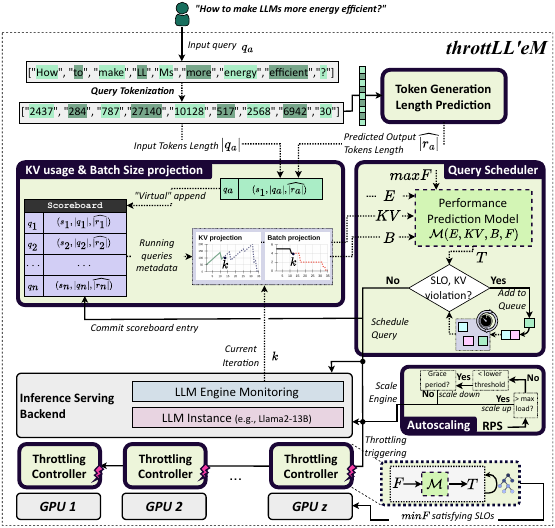}
        \caption{\textit{throttLL'eM} Overview}
        \label{fig:framework-overview}
\end{figure}

Building on insights from Sec.~\ref{sec:motivation}, we introduce \textit{throttLL'eM}, an energy-efficient, SLO-aware LLM inference serving system. 
\textit{throttLL'eM} prioritizes critical SLOs, namely E2E and TBT, delivering energy-efficient inference by dynamically adjusting GPU frequency to match workload demands.
Figure~\ref{fig:framework-overview} shows an overview of \textit{throttLL'eM}.
New queries are processed by the \textit{generation length prediction} component, which estimates the number of generated tokens.
This prediction enables accurate \textit{KV cache usage and batch size projections}, allowing the system to evaluate potential SLO violations.
If no violations are anticipated, the query is scheduled; otherwise, it is queued.
Finally, \textit{throttLL'eM} also includes an \textit{autoscaling mechanism} to dynamically adjust LLM instance parallelism based on workload, and a \textit{throttling controller} to optimize GPU frequency and energy consumption.
Table~\ref{tab:terminology} summarizes the terminology used in this section.
\begin{table}[b]
\caption{Terminology used in Section~\ref{sec:methodology}}
\label{tab:terminology}
\centering
\resizebox{\columnwidth}{!}{%
\begin{tabular}{c|l}

\Xhline{3\arrayrulewidth}
\rowcolor{gray!20}
\textbf{Term} & \textbf{Definition} \\

\Xhline{3\arrayrulewidth}

$s_i$ & The iteration at which query $q_i$ is scheduled\\ 
\rowcolor{gray!20}
$|q_i|$ & The input length of query $q_i$\\
$\widehat{|r_i|}$ & The predicted generation length of query $q_i$\\
\rowcolor{gray!20}
$KV$ & Vector of predicted KV cache usage for the LLM engine\\
$KV_{q_i}$ & Vector of predicted KV cache usage for query $q_i$\\
\rowcolor{gray!20}
$B$ & Vector of predicted Batch size per iteration \\
$T$ & Vector of predicted token generation throughput (IPS) \\
\rowcolor{gray!20}
$N$ & Number of tokens per KV cache block \\
$\widehat{t_R}(q_i)$ & The estimated remaining time for query $q_i$\\
\rowcolor{gray!20}
$t_{dead}(q_i)$ & The deadline for the E2E SLO of query $q_i$\\
$\mathcal{M}$ & The performance prediction model used by \textit{throttLL'eM} \\
\rowcolor{gray!20}
$\max{F}$ & The maximum frequency supported by the GPU\\
$E$ & The size of the engine (number of GPUs)\\

\Xhline{3\arrayrulewidth}
\end{tabular}}
\end{table}

\subsection{Generation Length Prediction}
\label{subsec:generation-length-prediction}
\checkednew{Accurate generation length prediction is essential for optimizing LLM inference, as it can mitigate KV cache overheads (cf. Sec.~\ref{subsec:motivation-kv-cache-analysis}) and prevent out-of-memory errors~\cite{jin2023s3} due to dynamic memory allocations~\cite{kwon2023pagedattention}.
Given the direct correlation between the number of generated tokens and allocated KV blocks, length prediction is crucial for preventing such issues. 
However, accurately predicting generation length is challenging due to the stochasticity of LLMs, the diversity of input queries, and other factors influencing output length.}

\checkednew{Recent works have proposed various approaches to tackle the aforementioned problem~\cite{jin2023s3, qiu2024efficient, hu2024inference, NEURIPS2023_ce7ff340}.
In~\cite{jin2023s3, hu2024inference}, the authors model length prediction as a classification task, fine-tuning DistilBERT~\cite{sanh2019distilbert} and OPT-125M~\cite{zhang2022opt} respectively, to classify output length into ``\textit{buckets}''.
In contrast, the authors of \cite{qiu2024efficient} use a regression model, appending a two-layer fully connected network to a BERT base model\cite{devlin2018bert}. 
In~\cite{NEURIPS2023_ce7ff340}, the authors propose the \textit{"Perception in Advance"} method, where the model estimates the length of its response before generating the output. All approaches report errors up to $\approx30$\%.
In this paper, we consider a prediction model that estimates the expected generation length ($\widehat{|r_a|}$) of each input query.
This predicted token generation length ($\widehat{|r_a|}$) is propagated to the KV cache usage \& Batch Size projection component.}

\subsection{KV usage \& Batch Size projection}
\label{subsec:KV_cache_projection}

The \textit{KV usage \& Batch Size projection} component generates vectors containing the projected KV cache utilization ($KV$) and batch size ($B$) at each future iteration until all currently scheduled requests terminate.
By assuming an oracle-based generation length predictor and considering the iterative nature of the token generation process, it is possible to create accurate projections for these values over time.
Specifically, since each request in the batch generates one token per iteration (each corresponding to a fixed KV cache block allocation size), we can predict the cumulative KV cache demands and batch sizes for upcoming iterations using an analytical model.

To keep track of scheduled requests, a \textit{Scoreboard} is kept, containing the following metadata for each query ($q_i$) in the batch: i) the iteration at which the query was scheduled ($s_i$), ii) the input length of the query ($|q_i|$) and iii) its predicted generation length ($\widehat{|r_i|}$). 
Based on these values, we can determine that a query will complete its generation process at iteration $s_i + \widehat{|r_i|}$, i.e., $\widehat{|r_i|}$ iterations after it initially scheduled.
By repeating this process for each entry in the \textit{Scoreboard}, we can identify the final iteration of all scheduled queries, allowing us to generate a vector $B=[B_1,B_2,...,B_j,...,B_n]$, where $B_j$ is the projected batch size at each future iteration $j\in[1,n]$ and $n$  is the iteration at which the last query will be completed.

A similar approach can be followed to calculate the \textit{KV cache} utilization for each iteration. 
At iteration $j$, request $q_i$ will have generated $j - s_i$ tokens\footnote{Under the constraint $s_i \le j < s_i +\widehat{|r_i|}$ which ensures that $q_i$ is scheduled.}. 
The KV cache also stores the precomputed values for the $|q_i|$ tokens that constitute the input of $q_i$. 
As each block has a capacity of $N$ tokens, where $N$ is a compile time parameter, the KV cache blocks allocated to request $q_i$ ($KV_{q_i}[j]$) are  given by Equation~\ref{equ:single_request_kv}.

\begin{equation}
\label{equ:single_request_kv}
KV_{q_i}[j] =
\begin{cases}
    \lceil (j - s_i + |q_i|) / N \rceil& \text{, } s_i \le j < (s_i + \widehat{|r_i|})\\
    0  & \text{, otherwise.}
\end{cases}
\end{equation}

\noindent  The total KV cache block usage of the inference engine at each future iteration $j$ can finally be calculated by summing all $KV_{q_i}$. This is the $KV$ of Figure~\ref{fig:framework-overview}, given by Equation~\ref{equ:global_kv}, where $KV = [KV_1,KV_2,...KV_j,...,KV_n]$.

\begin{equation}
\label{equ:global_kv}
KV[j] = \sum^{\forall q_i}{KV_{q_i}[j]}
\end{equation}

\noindent Last, when a query terminates, its entry is struck from the \textit{Scoreboard}, signalling the deallocation of all associated KV cache blocks. 
 The introduction of this component allows us to project the usage of the KV cache and the batch size into the future, under the assumption that no new requests will arrive. \AK{We measure an average latency of $<2$ ms for this component.} 

\noindent\textbf{Projections for new queries:} When a new query $q_a$ is submitted and before it is scheduled, its metadata are ``virtually" appended to the \textit{Scoreboard} with $s_a$ being set to the current iteration $k$, thus, allowing us to assess how future KV cache usage and batch size would be affected by the immediate scheduling of the request. 
If the query is indeed scheduled by \textit{throttLL'eM}'s scheduler (Sec.\ref{subsec:query-scheduler}), this ``virtual" change is committed to the \textit{Scoreboard}, otherwise, it is rolled back.

\subsection{LLM Query Scheduler}
\label{subsec:query-scheduler}
The \textit{Query Scheduler} is responsible for orchestrating the execution of incoming requests on the LLM engine.
To identify potential SLO violations, a \textit{Performance Prediction Model} is employed. 
If no SLO violations are predicted, the query is scheduled for execution; otherwise, it is placed in a waiting queue until it can be served without compromising any SLO.

\subsubsection{Performance Prediction Model}
\label{subsubsec:performance-prediction-model}
The \textit{Performance Prediction Model} ($\mathcal{M}$) predicts the throughput of the system in terms of iterations per second (IPS).
Building on the high correlation of throughput to KV cache usage, GPU frequency, and batch size, as demonstrated in Sec.~\ref{subsec:motivation-heatmaps} and Sec.~\ref{subsec:motivation-kv-cache-analysis}, the model leverages these metrics as input features to accurately forecast system performance at the iteration level.

\noindent\textbf{Training data collection:} Capturing the entire design space for training the predictive model is prohibitive due to i) the fine granularity of GPU frequencies (15 MHz steps); ii) the extensive range of possible values for KV block usage; and iii) the large batch sizes, which can reach up to 512 instances in few-parameter models with a high degree of parallelism, or on GPU infrastructures with higher memory capacity~\cite{agrawal2023sarathi}. 
To address this challenge, we employ a systematic sampling approach to acquire a representative subset of the design space.
Specifically, we employ a request generator that profiles the LLM inference engine across different batch sizes by spawning queries with predefined generation lengths, ensuring maximum utilization of the KV cache at the final iteration. 
For each batch size, the generator spawns a number of requests equal to the batch size to cover the entire KV cache usage range, \AK{ensuring that the edges of the profiling space are present in the dataset}. 
A monitoring agent is deployed to continuously log KV cache usage, GPU frequency, batch size, and the experienced IPS every second. 
\AK{During this process, the GPU frequency is randomly changed (for high frequency range coverage) after each measurement is completed and before starting a new one, in order to ensure that the frequency remains consistent for the duration of each measurement.}
This data collection process is repeated for all supported tensor parallelism (TP) levels, generating a dataset for each LLM that captures the relationship between the input variables and IPS.

\noindent\textbf{Model selection \& training:}  
\textit{throttLL'eM} adopts a machine learning (ML) approach to model the relationship between system metrics and IPS.
Given that all input features are upper and lower bounded for a given LLM, we employ a Gradient Boosted Decision Tree~\cite{friedman2001greedy,xgboost} as our ML model, which has been shown to provide accurate interpolation predictions for bounded datasets with multiple interacting features~\cite{malistov2019gradient}.
The model is lightweight ($\approx3$ ms inference latency on CPU), meeting the design constraint of low inference latency. 
This is vital not only for enabling iteration-level predictions but also because the model resides in the critical path of the \textit{Scheduler}.
We use the following features as input to the model: i) engine size; ii) batch size; iii) KV cache usage; and iv) GPU frequency, training it to predict the throughput (IPS).

\subsubsection{Query Scheduling \& Admission Control}
\label{subsub:using_performance_prediction}
Upon receiving a new query, the \textit{Scheduler} initiates an admission control process, using the projected batch ($B$) and KV cache size ($KV$) vectors, generated by the \textit{KV usage \& Batch Size projection} mechanism using the ``virtual'' \textit{Scoreboard}.
The \textit{Scheduler} must determine if the new query can be scheduled without causing any SLO violations. This involves three checks:

\noindent\textbf{1. KV cache assessment:} The projected KV cache utilization ($KV$) is compared to the available KV cache capacity of the inference engine.
If any value in $KV$ exceeds this capacity, the query is queued to prevent significant performance degradation from KV block swapping to main system memory.

\noindent\textbf{2. TBT SLO compliance:} For queries meeting KV cache constraints, the \textit{Scheduler} uses $\mathcal{M}$ to estimate throughput for each projected batch size ($B_j \in B$) and KV cache usage ($KV_j \in KV$) pair.
The model is queried with: i) the current size of the LLM inference engine, ii) projected batch sizes $B$, iii) projected KV cache utilization $KV$, and iv) the maximum available GPU frequency, reflecting peak theoretical performance. 
The model predicts a vector $T$ of maximum attainable throughput (IPS) per future iteration.
\AK{Since $\text{TBT} = 1 / \text{IPS}$}, the \textit{Scheduler} calculates the inverse values of $T$ to form a vector $T'$, representing TBT per iteration.
If the average value of $T'$ exceeds the TBT SLO, the request is queued.

\noindent\textbf{3. E2E SLO compliance:} Given the TBT vector $T'$, we can determine the estimated time to reach a future iteration $j$, by calculating the vector $\widehat{T_R} = [T_{R_1},T_{R_2},...T_{R_j},...,T_{R_n}]$, i.e., performing a cumulative sum on the values of $T'$ as follows:
\begin{equation}
\label{equ:estimated_remaining_time}
    \widehat{T_{R_j}} = \sum_{i=k}^{i \le j}{T'_i} = \sum_{i=k}^{i \le j }{T_i^{-1}}
\end{equation}
For request $q_i$, finishing at iteration $l = s_i + \widehat{|r_i|} - k$, the remaining time is given from element $T_{R_l}$. The current iteration $k$ is subtracted as future iterations are relative to $k$.
Thus, for each $q_i$ which is predicted to complete at iteration $l$, the \textit{current time} ($t_{cur}$) is a added to $T_{R_l}$ and compared to the \textit{deadline} imposed by the E2E SLO ($t_{dead}(q_i)$):
\begin{equation}
\label{equ:validate_SLO}
\forall q_i, T_{R_l} + t_{cur} < t_{dead}(q_i)
\end{equation}

For queued requests, this admission control process is repeated until the \textit{Scheduler} decides that they can be safely admitted to the batch without causing violations.
If all checks are successful, the newly arrived request is scheduled and any changes to the \textit{Scoreboard} are committed. 
Otherwise, the request is queued and changes are rolled back. 
If a new request cannot be scheduled without violation of its own E2E SLO, \AK{but does not cause violations}, it is scheduled but marked as ``lost", signaling the \textit{Scheduler} to ignore it during future SLO validations.

\subsection{LLM autoscaling mechanism}
\label{subsec:autoscaling}
Effective autoscaling is crucial for optimizing both performance and energy efficiency in LLM inference systems.
As shown in Sec.~\ref{subsec:motivation-parallelism}, tensor parallelism (TP) forms the most efficient approach for scaling the LLM engine to multiple GPUs on the same node.
To this end, \textit{throttLL'eM}'s autoscaling mechanism relies on managing the TP of the LLM engine to match the incoming request rate (RPS), ensuring that the system operates efficiently under varying loads.

Autoscaling is driven by a 10-second interval monitoring agent that determines the minimum engine size required to handle the current workload.
This decision is based on pre-characterized performance profiles, which define the maximum load each engine can sustain without introducing long tail latencies (cf. Sec.~\ref{subsec:experimental-setup}).
To mitigate the significant provisioning latency ($>$20s) of new inference servers, we employ ``\textit{shadow instancing}"\cite{shadowMarcusChow,dhakal2020gslice}, thus
masking switching overheads.
This involves a ``warm-up'' phase, where new requests are served by the current engine until the new one becomes operational. 
Afterwards, a ``transition'' phase occurs, during which the old engine gracefully shuts down, while the new engine takes over, serving all new requests.
During shadow instancing, both engines run simultaneously, consuming energy without providing peak performance capacity until fully loaded. 

\checkednew{To regulate autoscaling and minimize energy waste, we establish a policy where newly spawned engines are given a ``grace period'' equal to their spawn time. 
During this period, the autoscaler may switch to a larger engine to adapt to sudden load spikes but may not switch to a smaller engine, preventing premature downscaling and unnecessary shadow instancing.
The grace period is renewed each time the measured RPS is within the engine's workload constraints. 
This policy prioritizes adherence to SLOs by always allowing scale-up decisions while minimizing scale-downs and unnecessary shadow instancing, thereby reducing energy consumption.}

\subsection{GPU Frequency Throttling Controller}
\label{subsub:frequency_selection}
The throttling controller is triggered upon admission of a new query to the inference engine.
Its goal is to identify the minimum GPU frequency that adheres to the target SLOs, unlike the \textit{Scheduler}, which validates SLO compliance at maximum frequency. 
Since this component is invoked subsequent to successful SLO validation at maximum frequency, the existence of at least one SLO-compliant GPU frequency is guaranteed.
The controller receives the $KV$ and $B$ vectors from the \textit{KV and Batch projection} mechanism, performing a binary search over the GPU frequency range. This involves: 1) Recalculating $T$, $T'$, and $\widehat{T_R}$ for each frequency; and 2) Checking TBT and E2E SLO compliance using the same procedure as the \textit{Scheduler}.
This process determines the minimum SLO-satisfying frequency. The selected frequency is applied to all the GPUs serving the LLM instance, thus, minimizing operational complexity and mitigating barrier synchronization issues due to stragglers~\cite{yang2020mitigating}.
If a ``lost'' request is present in the current batch, the binary search is bypassed, and the maximum GPU frequency is selected in a best-effort attempt to meet the ``lost'' SLO. \AK{We measure the combined latency of the \textit{Scheduler} and \textit{GPU frequency throttling} components at 35 ms under heavy load.}

\subsection{Discussion}
\label{subsec:discussion}
\noindent\textbf{Addressing generation length prediction errors: } The effectiveness of \textit{throttLL'eM} relies on accurate generation length predictions, yet such errors are unavoidable. 
Queries terminating earlier than expected primarily impact energy savings, as their SLOs could have been met at lower frequencies. 
Conversely, queries exceeding the predicted length risk violating SLOs. 
To mitigate this, we conservatively adjust the predicted generation length ($\widehat{|r_i|}$) by a factor proportional to the error induced by the predictor.
If the actual length surpasses the adjusted $\widehat{|r_i|}$ during the generation process, the \textit{Scoreboard} entry for the request is updated to the maximum length limit (\texttt{\small max\_tokens}) supported by the model.

\noindent\textbf{\textit{throttLL'eM} and TTFT optimization:} \textit{throttLL'eM} does not explicitly optimize the \textit{prefill} phase, a compute-intensive part of LLM inference.
While the \textit{prefill} phase typically constitutes a smaller proportion of overall latency compared to the generation phase~\cite{patel2024splitwise}, it still represents a potential performance bottleneck.
Increasing GPU frequency for a single iteration (upon scheduling a new query) would be an inefficient way to manage TTFT due to the frequency switching overhead (avg. 200 ms) compared to the short \textit{prefill} phase duration (avg. $\approx$175 ms). The Splitwise framework~\cite{patel2024splitwise} offers a promising alternative by decoupling the \textit{prefill} and \textit{generation} stages onto separate instances, allowing for independent optimization.

\noindent\textbf{\textit{throttLL'eM} scalability:} The proposed solution uses TP to distribute LLMs across GPUs, an approach shared with other concurrent works~\cite{patel2024splitwise,patel2023polca}. 
TP offers low latency but requires high performance interconnects and may not scale beyond a single node. 
As discussed in~\cite{agrawal2023sarathi}, for larger models which need to be distributed across nodes, PP is often used for inter-node communication, while TP is employed intra-node to take advantage of its increased performance. 
In this case, the modelling methodology presented still applies within each node but additional components would be required in order to orchestrate execution across stages.
\textit{throttLL'eM} is orthogonal to DDP, used by~\cite{patel2023polca,patel2024splitwise} to deploy multiple LLM instances.

\noindent\textbf{\textit{throttLL'eM} overheads:} The proposed solution employs a generation length prediction model. Previous works~\cite{qiu2024efficient} show that such models introduce an average latency of $7.6$ms, which is negligible when compared to the multiple seconds of the inference process. 
throttLL'eM also requires profiling to generate the performance prediction model: a process that takes up to a day for LLaMa3-70B. 
However, this is a one-time cost that can be amortized over the lifetime of the service and across multiple instances.
The prediction model requires less than a second of training on a multi-core CPU, while inference takes on average $3$ms. Overall, when throttLL'eM is deployed, the \textit{Scheduler} and \textit{frequency throttling} components add $35$ms to i) forecast KV cache utilization and batch sizes ($2$ms), ii) check multiple frequencies and schedule a query.
Finally, setting a new frequency requires $8$ms per GPU, which however occurs after request scheduling, off the critical path.

\section{Evaluation}

\label{sec:eval}

\subsection{Experimental Setup}
\label{subsec:experimental-setup}

\noindent\textbf{HW/SW stack:} We deploy \textit{throttLL'eM} on a \texttt{\small p4d.24xlarge} AWS EC2 instance equipped with $8\times$ NVIDIA A100 40GB GPUs~\cite{aws-ec2-p4dspecs}.
For the software stack, we use NVIDIA's Triton server~\cite{triton-inference-server} and TensorRT-LLM backend~\cite{tensorRT-LLM-backend}
which provides high-performance LLM inference on NVIDIA GPUs.
We enable typical LLM performance optimizations, such as: i) inflight fused batching~\cite{gyeong2022orca}; ii) paged KV caching\cite{kwon2023pagedattention}; and iii) Flash Attention\cite{dao2022flashattention}.

\setlength{\extrarowheight}{1.4pt} %

\begin{table}[b]
\caption{Performance profiles of examined LLMs.}
\label{tab:performance-profiles}
\centering
\resizebox{.9\columnwidth}{!}{%
\begin{tabular}{c|c|c|c|c|}
\Xcline{2-5}{3\arrayrulewidth}
\textbf{}  & \begin{tabular}[c]{@{}c@{}}\textbf{TP}\\ \it{(\#)}\end{tabular}  & \begin{tabular}[c]{@{}c@{}}\textbf{max load}\\ \textit{(RPS)}\end{tabular} & \begin{tabular}[c]{@{}c@{}}\textbf{E2E SLO}\\ \textit{(p$99^{th}$)}\end{tabular} & \begin{tabular}[c]{@{}c@{}}\textbf{KV blocks}\\ \it{(\#)}\end{tabular} \\
\Xcline{1-5}{3\arrayrulewidth}
\multicolumn{1}{|l|}{\textbf{\small Llama3-8B}}  & 1  & 13.0 & 37.7 & 1033\\
\hline
\rowcolor[HTML]{EFEFEF}\multicolumn{1}{|l|}{\textbf{\small Llama2-13B}} & 1 & 1.125 & 22.7 & 120 \\
\hline
\multicolumn{1}{|l|}{\textbf{\small Llama2-13B}} & 2  & 4.0 & 30.2 & 439 \\
\hline
\rowcolor[HTML]{EFEFEF}\multicolumn{1}{|l|}{\textbf{\small Llama2-13B}} & 4 & 7.5 & 31.3 & 1050 \\
\hline
\multicolumn{1}{|l|}{\textbf{\small LLama3-70B}} & 8 & 7.0 & 44.0 & 2205 \\              \Xcline{1-5}{3\arrayrulewidth}
\end{tabular}}
\end{table}

\noindent\textbf{Examined LLMs and SLO definition:} We examine LLMs of different sizes and configurations from the LLaMa family~\cite{meta_llama}, presented in Table~\ref{tab:performance-profiles}.
The TBT SLO is set as to an average latency below 200 ms, corresponding to the average human reading speed of 250 words per minute~\cite{brysbaert2019many}, a value adopted as a target constraint by MLPerf~\cite{mlcommons-llama,mlperf}.
To define the E2E SLO, we profile the LLMs at maximum GPU frequency using the MLPerf benchmark~\cite{mlperf}, gradually increasing the RPS until reaching saturation, characterized by long tail latencies in response times. 
The E2E SLO is set to the p$99^{th}$ response time under maximum load.
Table~\ref{tab:performance-profiles} summarizes this information.

\noindent\textbf{Load generation:} Workload is generated using invocation traces from Azure~\cite{patel2024splitwise}, analyzed in Sec.~\ref{subsec:motivation-trace-analysis}.
Since the trace has a peak RPS of $\approx8.25$, we right-scale the invocation rate to match the maximum load of the respective evaluated engine (reported in Table~\ref{tab:performance-profiles}).
As the trace does not provide the actual context of input and output due to data protection regulations, we generate synthetic queries that match the prompt and generation lengths of each item in the dataset.

\subsection{Performance Prediction Model Evaluation}
\label{sub:perf_predict_eval}
Table~\ref{tab:modelling_accuracy} reports the $R^2$ \textit{score}, \textit{Mean Absolute Error} (\textit{MAE}) and \textit{Mean Absolute Percentage Error} (\textit{MAPE}) of the performance prediction model ($\mathcal{M}$) employed by \textit{throttLL'eM} (Sec.~\ref{subsubsec:performance-prediction-model}), for each one of the examined engines and under two different train/test splits.
Using a $90/10$ split, $\mathcal{M}$ accurately predicts target IPS with $R^2$ scores $\ge 0.97$, often exceeding $0.98$.
MAPE is kept below $5.8$\% for all configurations, reaching a low of $2.8$\% for Llama2-13B when running on a single GPU. 
MAE, providing a more tangible representation of the results, indicates mispredictions under $1$ IPS on average for all configurations. 
With a $10/90$ train/test split, prediction performance remains robust, featuring $R^2$ scores above $0.96$, MAE below $1.01$ IPS, and a MAPE increase of $0.7$\% at most. 
These results demonstrate that high accuracy is achievable even with sparse training sets.

\subsection{KV cache and Batch size Projection Mechanism Evaluation}
\label{sub:kv_eval_analysis}

\begin{figure}[t]
    \centering
    \begin{minipage}[t]{0.7\columnwidth}
        \vspace{-10pt}
        \centering
        \subfloat[Batch Size.]{
            \includegraphics[width=0.3\textwidth]{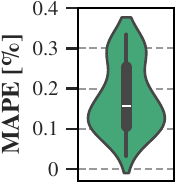}
            \label{fig:eval-batch-mape}
        }
        \subfloat[KV usage.]{
            \includegraphics[width=0.27\textwidth]{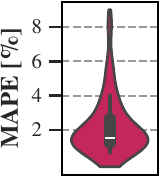}
            \label{fig:eval-kv-mape}
        }
        \subfloat[Drift.]{
            \includegraphics[width=0.275\textwidth]{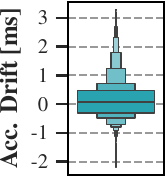}
            \label{fig:eval-drift}
        }
    \end{minipage}
    \begin{minipage}[t]{0.27\columnwidth}
        \centering
        \vspace{0pt}
        \begin{tabular}{|c|c|}
                    \Xcline{1-2}{3\arrayrulewidth}
                    \multicolumn{2}{|c|}{\textbf{\small Avg. errors}} \\
                    \Xcline{1-2}{3\arrayrulewidth}
                    \small \textbf{Batch \%} & \small 0.19  \\
                    \hline
                    \rowcolor[HTML]{EFEFEF}\small \textbf{KV \%} & \small 2.26 \\
                    \hline
                    \small \textbf{Drift ms} & \small 0.43 \\
                    \Xcline{1-2}{3\arrayrulewidth}
                \end{tabular}
    \end{minipage}
    \caption{Distributions and average values of various metrics}
    \label{fig:eval-projection}
\end{figure}

We evaluate the \textit{KV cache usage and batch size projection} component (Sec.~\ref{subsec:KV_cache_projection}) using custom generated micro-traces.
Each micro-trace sets a different GPU frequency and includes queries with random prompt and generation lengths.
We spawn all queries simultaneously \AK{which trigger \textit{throttLL'eM}'s projection and prediction mechanisms, generating vectors $KV$, $B$ and $\widehat{T_R}$ for the specified frequency, which contains the estimated remaining time to reach all future iterations. 
At the same time, the performance monitoring endpoint of Triton is continuously polled, logging the actual batch size, KV cache usage and arrival time at each iteration. 
Figures~\ref{fig:eval-batch-mape},~\ref{fig:eval-kv-mape} show the error distributions between actual and predicted batch size and KV cache usage respectively, while Fig.~\ref{fig:eval-drift} plots the accumulated drift, i.e., how much predicted times (elements of $\widehat{T_R}$) diverge from actual arrival times, divided by the number of elapsed iterations. 
Overall, batch size projection errors are kept to a minimum, averaging an error of 0.19\%. The KV projection error is slightly higher while still remaining low at 2.26\% on average. These small variations are attributed to the latency experienced between request submission and admission, during which execution progresses, occasionally resulting in off-by-one errors when calculating the final iteration of each request. Finally, each LLM iteration introduces a relatively small average drift of 0.43 ms, compared to the TBT (15 to 30 ms) for the same inference engine, shown in Fig.~\ref{fig:motivation-tbt}.} This drift is a combined effect of i) the aforementioned errors in KV cache usage and Batch size projections; and ii) estimation errors of the performance prediction model $\mathcal{M}$.

\setlength{\extrarowheight}{1.4pt} %

\begin{table}[b]
\caption{Performance Prediction Model Evaluation}
\label{tab:modelling_accuracy}
\centering
\resizebox{\columnwidth}{!}{%
\begin{tabular}{l|c|c|c||c|c|c|}
\Xcline{2-7}{3\arrayrulewidth}
& \multicolumn{3}{c||}{\textbf{\textit{train = 90\%}}}                                                                         & \multicolumn{3}{c|}{\textbf{\textit{train = 10\%}}} \\ \Xcline{2-7}{3\arrayrulewidth}

              & \begin{tabular}[c]{@{}c@{}}\textbf{R\textsuperscript{2}}\\ \textit{\scriptsize (-)} \end{tabular} & \begin{tabular}[c]{@{}c@{}}\textbf{MAPE}\\ \textit{\scriptsize (\%)} \end{tabular} & \begin{tabular}[c]{@{}c@{}}\textbf{MAE}\\ \textit{\scriptsize (iters/s)}\end{tabular} 
              & \begin{tabular}[c]{@{}c@{}}\textbf{R\textsuperscript{2}}\\ \textit{\scriptsize (-)} \end{tabular} & \begin{tabular}[c]{@{}c@{}}\textbf{MAPE}\\ \textit{\scriptsize (\%)} \end{tabular} & \begin{tabular}[c]{@{}c@{}}\textbf{MAE}\\ \textit{\scriptsize (iters/s)}\end{tabular} \\ \cline{2-7} 
\hline
\multicolumn{1}{|l|}{\textbf{Llama3-8B-TP1}} & 0.99 & 4.1 & 0.85 & 0.98 & 4.2 & 0.93 \\
\hline
\rowcolor[HTML]{EFEFEF}\multicolumn{1}{|l|}{\textbf{Llama2-13B-TP1}} & 0.98 & 2.8 & 0.74 & 0.97 & 3.0 & 0.79 \\
\hline
\multicolumn{1}{|l|}{\textbf{Llama2-13B-TP2}} & 0.99 & 3.1 & 0.90 & 0.99 & 3.4 & 0.99 \\
\hline
\rowcolor[HTML]{EFEFEF}\multicolumn{1}{|l|}{\textbf{Llama2-13B-TP4}} & 0.99 & 3.3 & 0.97 & 0.99 & 3.4 & 1.01 \\
\hline
\multicolumn{1}{|l|}{\textbf{Llama3-70B-TP8}} & 0.97 & 5.8 & 0.69 & 0.96 & 6.5 & 0.77 \\
\Xhline{3\arrayrulewidth}
\end{tabular}}
\end{table}

\subsection{\textit{throttLL'eM} end-to-end Evaluation}
\label{sub:solution_eval}

\subsubsection{\textit{throttLL'eM without autoscaling}}
We first evaluate \textit{throttLL'eM} without autoscaling, under the assumption of i) an oracle generation length predictor (predicted length $\widehat{|r_a|}$ equals actual generation length) and ii) error-prone predictors with 15\% and 30\% p$95^{th}$ error, reflecting typical prediction errors reported in the literature~\cite{jin2023s3,hu2024inference,qiu2024efficient}.
To simulate prediction errors for the known-length queries of the Azure trace, we add Gaussian noise to the reported lengths using $\sigma$ values corresponding to 15\% and 30\% p$95^{th}$ prediction errors.

\begin{figure}[t]
\centering
\subfloat[End-to-End latency distribution. The red dotted line denotes the E2E SLO of each model as defined in Table~\ref{tab:performance-profiles}. Text boxes report the $99^{th}$ E2E latency percentile achieved by the different approaches.]{
    \includegraphics[width=\columnwidth]{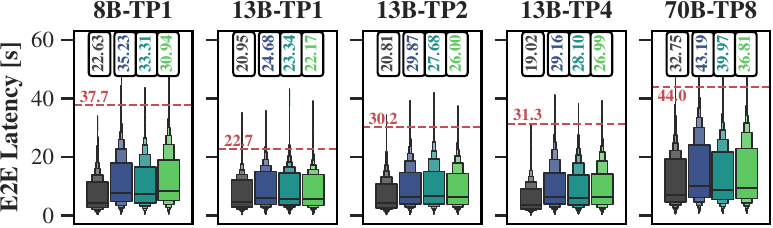}
    \label{fig:eval-error0-e2e}
}

\subfloat[TBT distribution. The red dotted line denotes the TBT SLO ($200$ ms average latency between token generation). Text boxes report the average TBT latency achieved by the different approaches.]{
    \includegraphics[width=\columnwidth]{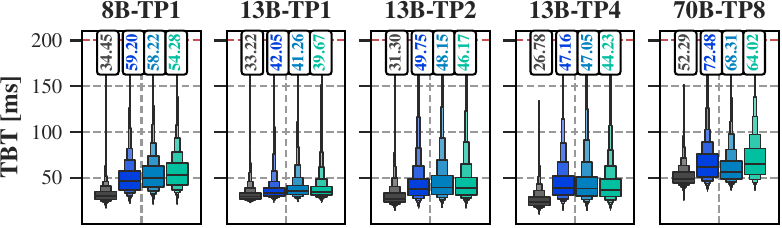}
    \label{fig:eval-error0-tbt}
}

\subfloat[Power Distribution (log scale). Text boxes report the energy efficiency (mean TPJ) achieved by the different approaches.]{
    \includegraphics[width=\columnwidth]{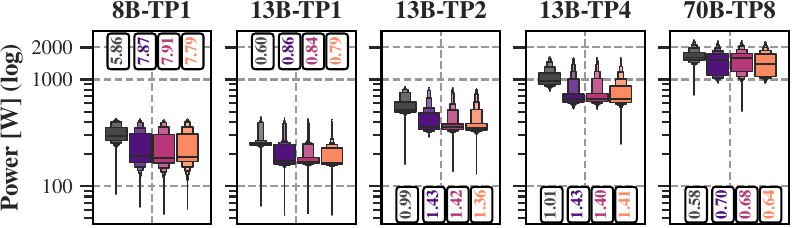}
    \label{fig:eval-error0-power}
}

\caption{Comparison of E2E, TBT and Power consumption between Triton~\cite{triton-inference-server} and \textit{throttLL'eM}. At each plot, the leftmost box represents Triton, followed by \textit{throttLL'eM} with error levels of $0\%$, $15\%$, and $30\%$, respectively.}\label{fig:main-analysis}
\end{figure}

Figure~\ref{fig:eval-error0-e2e} shows the distribution of E2E latencies for different engines when Triton\cite{triton-inference-server} and \textit{throttLL'eM} are deployed on the scaled trace. In almost all cases, with the exception of llama2-13b-TP1, \textit{throttLL'eM} achieves the specified $p99th$ latency SLO, being on average 1.45 seconds under the deadline. 
We attribute the inability of \textit{throttLL'eM} to achieve the E2E SLO for llama2-13b-TP1 to two factors. First, even the default Triton implementation stays just 1.7 seconds below the SLO, hinting that \textit{throttling} may not be applicable to this situation. 
Secondly, the very limited number of KV blocks available (120 per Tab.~\ref{tab:performance-profiles}) means that i) the engine is limited to lower batch size and throughput and ii) requests are often queued to avoid swapping to system memory. 
The TBT SLO is achieved in all cases, as depicted in Fig.~\ref{fig:eval-error0-tbt}, with the vast majority of TBT values falling well below 200 ms. Outlier TBT values can be attributed to \textit{inflight batching} overheads, as token generation stalls until the prefill phase of a new request is completed.

\textit{throttLL'eM} aims to increase the energy efficiency of LLM serving. Figure~\ref{fig:eval-error0-power} examines the distribution of power consumption values encountered by Triton and \textit{throttLL'eM} under different prediction error levels. Triton displays consistently high power draw for all LLM engines. On the other hand, the proposed solution offers reduced power consumption due to lower GPU frequencies. As \textit{throttLL'eM} dynamically adjusts frequencies to meet SLOs, a wide range of power values is encountered, including peak power equal to that of Triton when under high system pressure. Overall, \textit{throttLL'eM} offers 36.3\% higher energy efficiency on average with oracle predictions, dropping to 30.0\% at 30\% prediction error level. The highest increase in energy efficiency was experienced for llama2-13b-TP2, producing up to 44.3\% more tokens per Joule than Triton. Overall, energy consumption is reduced by an average of 24.7\% and up to 30.7\%.  

\begin{figure}[t]
\centering
    \includegraphics[width=\columnwidth]{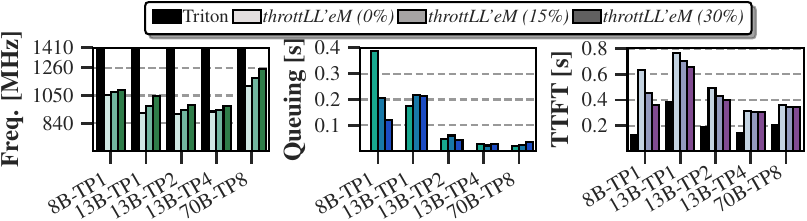}
\caption{Average GPU frequencies, Queue times and TTFT for various LLMs and configurations.}
\label{fig:additional-metrics}
\end{figure}

These improvements are enabled by the selection of lower frequencies. The barplots of Figure~\ref{fig:additional-metrics}, show that \textit{throttLL'eM} selects lower, more efficient frequencies than the default 1410 MHz, ranging between 950 and 1260 MHz on average, depending on the LLM engine and the prediction error of the predictor. Higher prediction errors lead to the selection of higher frequencies, as \textit{throttLL'eM} increases the predicted generation length, leading to frequency overprovisioning.

Finally, we analyze the overheads imposed by the introduction of our framework. \textit{throttLL'eM} maintains a queue where queries are held until the SLOs are validated. This introduces additional latency due to the time needed to perform the validation and the possible re-queuing of a request if it is predicted to cause violations. Figure~\ref{fig:additional-metrics} contains representative examples for both situations. Llama3-8b-TP1 exhibits relatively high queuing delays as high request arrival rates stress the \textit{Scheduler}, which can only schedule one request at a time. Llama2-13b-TP1 also exhibits delays, as the limited KV capacity leads to queuing until sufficient blocks become available. From the user's perspective queuing delays add to the latency of the prefill phase (TTFT). Figure~\ref{fig:additional-metrics} shows that \textit{throttLL'eM} increases TTFT when compared to Triton. A significant part of this increase is attributed to queuing, while the rest is due to the lower frequencies selected, which slow down the compute-bound prefill phase~\cite{agrawal2023sarathi}.

\subsubsection{\textit{throttLL'eM with autoscaling}}
Last, we evaluate \textit{throttLL'eM} including autoscaling using Llama2-13B's TP1, TP2 and TP4 configurations.
To ensure that all configurations are encountered, we scale the RPS range to $[0.75,7.5]$ (up to the max load of TP4) by applying different scaling factors to different areas of the trace, thus keeping its shape, but amplifying variations between highest and lowest RPS. 

\begin{figure}[t]
\centering
\subfloat[$p99th$ E2E. The red line repre-\\sents the latency threshold.]{
    \includegraphics[width=.47\columnwidth]{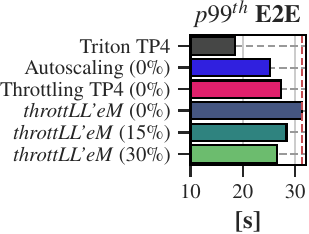}
    \label{fig:autoscale-e2e}
}
\subfloat[Energy\\consumption.]{
    \includegraphics[width=.2\columnwidth]{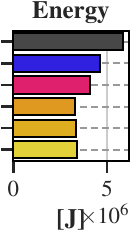}
    \label{fig:autoscale-nrg}
}
\subfloat[Energy\\Efficiency.]{
    \includegraphics[width=.2\columnwidth]{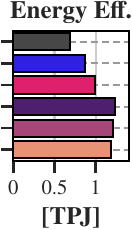}
    \label{fig:autoscale-tpj}
}

\caption{Comparison of various implementations on the trace.}\label{fig:main-subanalysis}
\end{figure}

To quantify the improvements afforded by both throttling and autoscaling, we compare \textit{throttLL'eM} to: i) the default Triton implementation on the TP4 engine; ii) Triton equipped with \textit{throttLL'eM} autoscaling; iii) \textit{throttLL'eM} without autoscaling (TP4 only) and oracle predictions. These cover the entire range of autoscaling and throttling combinations. Our solution is again evaluated using multiple prediction error levels. Figures~\ref{fig:autoscale-e2e} and \ref{fig:autoscale-nrg} show the E2E latency and energy consumption measured during the execution of the trace for all implementations. As expected, Triton features the lowest E2E latency but also the highest energy consumption. Enabling autoscaling (throttling) increases E2E latency as some performance is sacrificed for a 20.8\% (30.6\%) reduction in energy consumption. Using both, \textit{throttLL'eM} achieves the lowest energy consumption, 43.8\% (41.7\%) lower than Triton at 0\% (30\%) error level, but also the highest E2E latency, which however remains below the p$99^{th}$ threshold value for the TP4 engine. TBT SLOs are met in all cases (avg. TBTs $<$50ms, not shown). From an energy efficiency standpoint (Fig.~\ref{fig:autoscale-tpj}), Triton on TP4 achieves 0.69 TPJ while autoscaling-only and throttling-only achieve 0.87 and 0.99 TPJ respectively. \textit{throttLL'eM} which implements both achieves 1.19 to 1.23 TPJ depending on the error of the predictor, a 1.71$\times$ to 1.78$\times$ increase in efficiency over the baseline implementation. 

\subsection{\textit{throttLL'eM} Runtime Analysis}

\begin{figure}[t]
\centering   
    \includegraphics[width=\columnwidth]{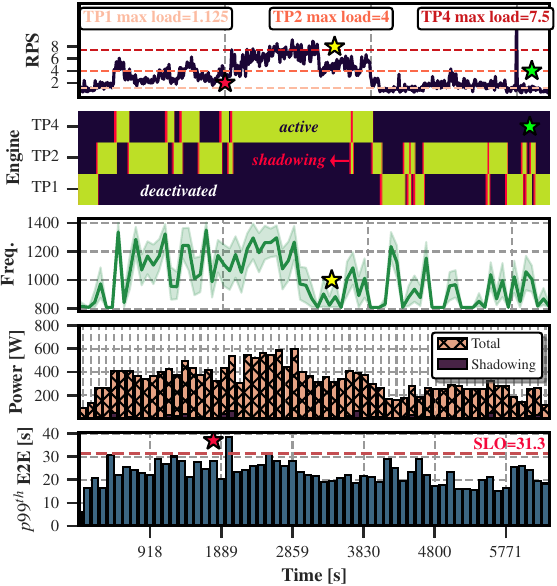}
    \caption{Runtime Analysis of \textit{throttLL'eM} on the scaled trace.}
    \label{fig:autoscaling-runtime}

\end{figure}

This section analyzes how \textit{throttLL'eM} adapts under varying load. Figure~\ref{fig:autoscaling-runtime} shows: i) experienced RPS; ii) the state of different TP engines; iii) applied frequencies; iv) average power draw; and v) p$99^{th}$ E2E latency over time. Different colors in the \textit{Engine} plot denote various LLM engine states. Average power is shown using hatched orange bars, while the power draw attributed to shadowing is shown in solid purple.

In general, \textit{throttLL'eM} selects lower frequencies when the \textit{active} engine is under light load. When however approaching the maximum load of an active engine (max load in Table~\ref{tab:performance-profiles}), frequency is increased above the energy efficiency \textit{``sweet spot''} to prevent SLO violations.

Breaking down the experienced p$99^{th}$ E2E latency to individual sections, it becomes evident that SLO violations may only be encountered when i) operating above the maximum load of the largest (TP4) engine or ii) during sudden increases in RPS, until up-scaling is completed. The former exceeds the rated capacity of the system, while the latter cannot be resolved unless startup times are reduced. Despite these transient violations, \textit{throttLL'eM} manages to achieve the p$99^{th}$ E2E SLO over the full duration of the trace.

Figure~\ref{fig:autoscaling-runtime} features some points of interest. The red stars mark transient E2E SLO violations due to a sudden load increase that cannot be absorbed by the TP2 engine while TP4 comes online. Yellow stars highlight how frequency is reduced when the instance is operating at lower load but cannot be downsized. Finally, green stars mark consecutive switches, as the encountered load remains close to the TP1/TP2 engine switching load boundary of 1.125 RPS.

\noindent$\star$ \textit{\textbf{Takeaways:} Autoscaling performs coarse-grained optimization by right-sizing system capacity. GPU throttling performs finer adjustments that cannot be achieved by autoscaling.}

\begin{figure}[t]
\centering   
    \includegraphics[width=\columnwidth]{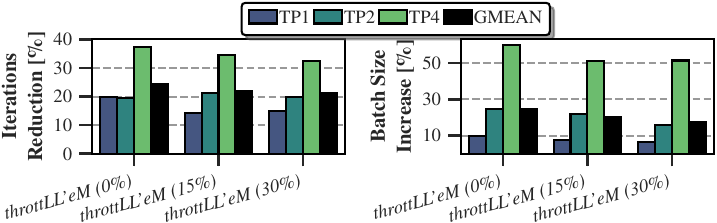}
    \caption{\textbf{a}) Reduction in iterations and \textbf{b}) Increase in avg. batch size of throttLL'eM to autoscale-only for all error levels.}
    \label{fig:result-interpretation}
\end{figure}

\begin{figure}[t]
\centering   
    \includegraphics[width=\columnwidth]{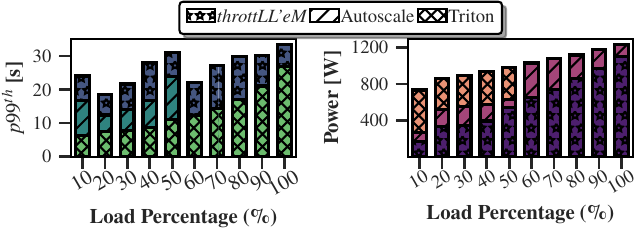}
    \caption{Comparison of \textbf{a}) $p99^{th}$ E2E and \textbf{b}) avg. Power for different implementations at increasing system load levels.}
    \label{fig:load-levels}

\end{figure}

\subsection{Extended Evaluation \& Comparison}

\subsubsection{Sources of Energy Gains} The observed energy gains in \textit{throttLL'eM} are driven by two key mechanisms. 
First, autoscaling adjusts the allocated capacity, resulting in a 20.8\% reduction in energy consumption compared to Triton TP4, as fewer GPUs are required, lowering static power draw across the entire trace.
Second, frequency scaling further reduces energy consumption by operating GPUs at lower frequencies, as shown in Sec.~\ref{subsec:motivation-heatmaps}. 
A side effect of this is that lower frequencies slow down the inference process, allowing more requests to accumulate in each batch, thereby artificially increasing batch size.
This improves energy efficiency (cf. Fig.~\ref{fig:motivation-nrgeff}), but also decreases the number of performed LLM iterations, leveraging the higher batch-level parallelism to generate more tokens per forward pass. 
Figure~\ref{fig:result-interpretation} illustrates this behaviour, showing how \textit{throttLL'eM} increases batch size and reduces iterations compared to autoscaling alone.
At 0\% predictor error level, \textit{throttLL'eM} manages to increase batch size by a geo-mean of $24.5\%$ across all engines, performing $24.3\%$ less iterations. 
Although higher errors result in higher frequencies, which diminish these benefits, the overall trend remains consistent.

\subsubsection{\textit{throttLL'eM} operation under varying loads} We examine the trade-off between energy efficiency and latency by progressively increasing the system load from 10\% to 100\% across different implementations.
Figure~\ref{fig:load-levels}b demonstrates how \textit{throttLL'eM} is able to reduce average power compared to Triton and autoscaling-only by downsizing engines and reducing frequencies. In this manner, \textit{throttLL'eM} trades-off latency, which as shown in Fig.~\ref{fig:load-levels}a, is brought close to that of peak load ($31.3$s per Tab.~\ref{tab:performance-profiles}). While autoscaling-only can significantly reduce the average power draw, e.g., at 10\% load level, in other cases the frequency scaling logic of \textit{throttLL'eM} offers additional gains which cannot be achieved with coarse-grained optimizations like autoscaling.

\begin{figure}[t]
\centering   
    \includegraphics[width=\columnwidth]{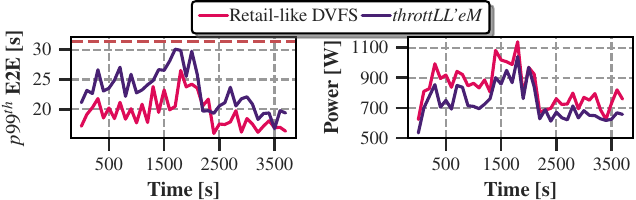}
    \caption{\textbf{a}) $p99^{th}$ E2E and \textbf{b}) Avg. power draw of throttLL'eM (w/o autoscaling) and a Retail-inspired implementation.}
    \label{fig:comparison}
\end{figure}
\subsubsection{Comparison with other DVFS schemes}
To demonstrate the effectiveness of our LLM-aware design choices, we compare \textit{throttLL'eM} with a SotA microservice-based DVFS scheme, similar to  ReTail~\cite{retail}.
ReTail automatically identifies input and application features that strongly correlate to E2E latency, in our case: frequency, batch size, KV cache utilization and input/output length. 
Using these, it trains a \textit{linear regression} model to predict latency and selects the lowest frequency that satisfies the SLOs for all requests in the queue.
Upon receiving a new request, ReTail evaluates available frequencies and chooses the minimum that meets SLO requirements.

To ensure a fair comparison, we used the LLaMa2-13B model in its TP4 configuration, as ReTail does not support autoscaling. Both implementations were tested at a 30\% predictor error level. 
Figure~\ref{fig:comparison}b shows that \textit{throttLL'eM} consistently achieves lower power consumption, while Fig.~\ref{fig:comparison}a demonstrates its ability to more aggressively approach the latency deadlines. 
We attribute this to i) \textit{throttLL'eM}’s capability to forecast KV cache utilization and batch size, and ii) a more accurate ML-based performance prediction model. 
In detail, ReTail relies on the currently known values of KV cache utilization and batch size for latency prediction, which however change over time, whereas \textit{throttLL'eM} uses forecasted values for future iterations.
Moreover, our analysis showed that ReTail's linear regression model struggles with E2E latency prediction, achieving $R^2$ close to 0.4. This is due to the non-linear E2E latency slowdown, as token generation proceeds. \textit{throttLL’eM} addresses this by using iteration-level predictions which are then aggregated to produce accurate E2E latency predictions.
Overall, across the entire trace throttLL'eM consumes $11.9$\% lower energy, a value that would have been higher had autoscaling  also been enabled.

\subsubsection{Evaluation on different systems/LLMs} 
\begin{figure}[t]
\centering   
    \includegraphics[width=\columnwidth]{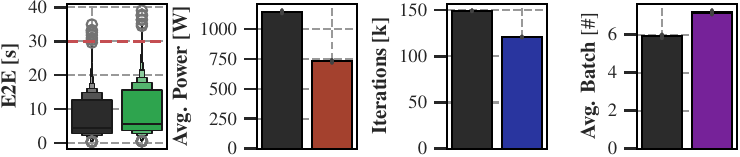}
    \caption{\textbf{a}) E2E latency \textbf{b}) Total Energy Consumption \textbf{c}) Total iterations and \textbf{d}) Average Batch size of Triton (left) and \textit{throttLL'eM} (right) of Mixtral 8x7B on NVIDIA A10G (TP8).}
    \label{fig:mixtral-a10g}
\end{figure}

We further examine how \textit{throttLL'eM} adapts to diverse systems and LLMs, by deploying a Mixtral 8x7B Mixture-of-Experts LLM~\cite{jiang2024mixtral} on a \texttt{g5.48xlarge} AWS instance with $8\times$ NVIDIA A10G 24GB GPUs. This system features lower performance than the previously examined one, lacking the HBM2 memory and the 600 GB/s NVLink interconnects of \texttt{p4d.24xlarge}. After profiling and training, $\mathcal{M}$ features an $R^2$ (MAPE) score of $0.98$ ($5.8\%$), displaying accurate modelling performance.  We scale the trace at the peak load of 1.25 RPS and a target latency of 30s.
Figure~\ref{fig:mixtral-a10g} displays the $p99^{th}$ E2E latencies and the average power draw of both implementations when serving the scaled trace. In this setup, \textit{throttLL'eM} is able to reduce average power by $35.9\%$, while still featuring a $p99^{th}$ E2E latency of $28.4$s, being well within the $30$s SLO for this configuration.

\section{Related Work}

\checkednew{LLM serving optimization is an active research topic, with many works aiming to improve the inference process~\cite{zhou2024survey}.}

\noindent\checkednew{\textbf{LLM inference optimization:} Recent research has explored various optimization techniques for LLM inference.
These efforts address different facets of LLM inference, including
i) efficient request batching~\cite{fang2021turbotransformers,gyeong2022orca} with \textit{inflight batching} introduced by Orca~\cite{gyeong2022orca} being a pioneering work in the field, ii) efficient LLM serving and request scheduling~\cite{gyeong2022orca,li2023alpaserve,agrawal2023sarathi,sheng2024fairness,spotserve,wu2023fast,hu2024inference,patel2024splitwise,zhong2024distserve,song2023powerinferfastlargelanguage, triton-inference-server,wang2023tabi} with NVIDIA's Triton~\cite{triton-inference-server} being a state-of-the-art solution for inference serving offering optimized performance, iii) LLM memory optimization~\cite{sheng2023flexgen,kwon2023pagedattention,fang2021turbotransformers,zhang2024h2o} with \textit{paged attention}~\cite{kwon2023pagedattention} being a notable innovation in the area, iv) model scaling and parallelism on multiple GPUs~\cite{aminabadi2022deepspeed,pope2023efficiently,jiand2024megascale}, v) speculative decoding and sampling~\cite{xia2024unlocking, miao2024specinfer,stern2018blockwise,chen2023accelerating, leviathan2023fast, xia-etal-2023-speculative}, vi) parameter sharing~\cite{zhou2022pets} and vii) optimization of GPU kernels~\cite{wang-etal-2021-lightseq,zhai2023bytetransformer,dao2022flashattention,dao2023flashattention,faster_transformer,aminabadi2022deepspeed} with Flash Attention~\cite{dao2022flashattention,dao2023flashattention} and NVIDIA's FasterTransformer~\cite{faster_transformer} being the most notable works. 
The work presented in~\cite{SoCC-LLM-queues} manages multiple queues for interactive and batched inference requests, avoiding potential head-of-line blocking.
While these works primarily target performance optimization, neglecting energy efficiency, they offer orthogonal techniques that can complement \textit{throttLL'eM}. 
Notably, as mentioned in Sec.~\ref{subsec:experimental-setup}, \textit{throttLL'eM} is built on top of Triton~\cite{triton-inference-server} which integrates optimizations like \textit{inflight batching}~\cite{gyeong2022orca}, \textit{paged attention}~\cite{kwon2023pagedattention} and \textit{flash attention}~\cite{dao2023flashattention}.}

\noindent\checkednew{\textbf{Power and Energy Efficient LLM inference serving:} Recent works explore the energy and power implications of LLM inference. Studies such as ~\cite{samsi2023words,stojkovic2024towards,chien2023reducing} characterize energy consumption and carbon emissions but lack frameworks for energy-efficient LLM inference. 
The authors of~\cite{wilkins2024offline} model the performance and energy consumption of LLM queries using an analytical approach.
They use this model to solve an offline scheduling optimization problem.
However, they overlook the dynamic nature of inference systems and KV cache implications (cf. Sec.~\ref{subsec:motivation-kv-cache-analysis}). 
In~\cite{kakolyris-CAL} a framework for energy-efficient LLM inference is proposed, but is limited to fixed-length queries (\texttt{\small max\_tokens}) on single-GPU models.
A concurrent work, \textit{\textmu-Serve}~\cite{u-serve-atc} performs intelligent model partition placement and introduces a multiplicative-increase-additive-decrease (MIAD)-based DVFS scheme to reduce power consumption while meeting the SLOs of LLM requests based on their predicted generation length and SLO slack.
Lastly, \cite{patel2023polca} presents POLCA, a framework for power oversubscription in LLM clusters. \textit{throttLL'eM} can complement POLCA by performing frequency scaling under power-capping constraints, enhancing the margins for power oversubscription.}

\section{Conclusion}

LLM inference presents an emerging workload with considerable energy footprint. In this work we performed an extensive study on the factors that affect the inference process, gathering insights on performance and energy efficiency. Guided by these insights, we designed \textit{throttLL'eM}, a framework that uses a predictive, ML-driven approach to increase the energy efficiency of LLM inference serving under SLOs. \textit{throttLL'eM} improves efficiency by over $1.71\times$, consuming up to $43.8$\% less energy than Triton.

 \section*{Acknowledgements}
The authors would like to thank the National Infrastructure for Research and Technology Network (GRNET) for funding the cloud infrastructure used to evaluate \textit{throttLL'eM}. They further acknowledge Georgios Zervakis and Panagiotis Hadjidoukas for granting access to the DGX Deep Learning System of the Computer Engineering and Informatics Department (CEID), University of Patras, utilized during development. Lastly, they are grateful to Konstantinos Kanellopoulos and the anonymous reviewers for their constructive feedback, which enhanced the quality of this paper.

This work has been partially funded by the PRIVATEER project. PRIVATEER has received funding from the Smart Networks and Services Joint Undertaking (SNS JU) under the European Union’s Horizon Europe research and innovation
programme under Grant Agreement No. 101096110. Views and opinions expressed are however those of the author(s) only and do not necessarily reflect those of the European Union
or SNS JU. Neither the European Union nor the granting authority can be held responsible for them.

\bibliographystyle{IEEEtranS}
\bibliography{references}

\end{document}